\documentclass[10pt]{bmc_article}    
\usepackage{epsfig}
\usepackage{cite} 
\usepackage{url}  
\usepackage{ifthen}  
\usepackage{multicol}   
\usepackage[utf8]{inputenc} 
\usepackage{bm}        
\usepackage{amssymb}   
\usepackage{mathrsfs}
\usepackage{amsmath,amsthm}

\def\X{\mathscr{X}}
\def\R{\mathbb{R}}
\usepackage{bm}
\newcommand{\QED}{\nobreak \ifvmode \relax \else
      \ifdim\lastskip<1.5em \hskip-\lastskip
      \hskip1.5em plus0em minus0.5em \fi \nobreak
      \vrule height0.75em width0.5em depth0.25em\fi}
\def\X{\mathscr{X}}
\def\R{\mathbb{R}}

\def\includegraphics{}
\newboolean{publ}
\newenvironment{bmcformat}{\setboolean{publ}{true}}{}
\begin{document}
\begin{bmcformat}
\title{
Classical Information--Theoretical View of Physical Measurements
and Generalized Uncertainty Relations
}
\author{Yoshimasa Kurihara\correspondingauthor%
         \email{Yoshimasa Kurihara\correspondingauthor - yoshimasa.kurihara@kek.jp}
      }
\address{%
Institute of Particle and Nuclear Studies, The High Energy Accelerator Organization (KEK), Tsukuba, Ibaraki 305-0801, Japan
}%

\maketitle
\begin{abstract}
Uncertainty relations are discussed in detail not only for free particles but also for bound states within the framework of classical information theory. Uncertainty relation for simultaneous measurements of two physical observables is defined in this framework for generalized dynamic systems governed by a Sturm--Liouville-type equation of motion. In the first step, the reduction of Kennard--Robertson type uncertainties because of boundary conditions with a mean-square error is discussed quantitatively with reference to the information entropy. Several concrete examples of generalized uncertainty relations are given. Then, by considering disturbance effects, a universally valid uncertainty relation is investigated for the generalized equation of motion with a certain boundary condition. Necessary conditions for violation (reduction) of the Heisenberg-type uncertainty relation are discussed in detail. The reduction of the generalized uncertainty relation because of the boundary condition is discussed by reanalyzing experimental data for measured electron densities in a hydrogen molecule encapsulated in a fullerene ${\rm C}_{60}$ cage.
\end{abstract}
\ifthenelse{\boolean{publ}}{\begin{multicols}{2}}{}
%
%
%
\section*{Introduction}
An important and interesting topic in quantum mechanics is interpretation of the uncertainty relation, which was first expounded by Heisenberg\cite{Heisenberg}. He introduced his uncertainty relation as a principle of quantum mechanics through a Gedanken experiment regarding position and momentum measurements of a point particle using an imaginary gamma-ray microscope. Kennard\cite{Kennard} and Robertson\cite{Robertson} generalized Heisenberg's uncertainty relation and proved it mathematically as a relation between standard deviations (square roots of variances) associated with noncommuting operator pairs. Recently, technical terms related to the uncertainty relation, such as variance of distributions, mean-square measurement errors, and disturbances due to measurements, have been clearly understood\cite{Ozawa1988zz,Ishikawa}, and a universally valid uncertainty relation has been obtained\cite{Masanao200321,PhysRevA.67.042105,Masanao2004350}. It is still necessary we pursue a better understanding of uncertainty relations and their applications. 
Experiments claiming to demonstrate a violation of the Heisenberg-type uncertainty relation\cite{10228899} must be carefully considered from various points of view\cite{Kurihara:arXiv1201.5151}.\\
\indent 
In this study, we present a universal view of physical measurements on the basis of the classical information theory. Here "classical" means that aspects such as commutation relations, Hilbert spaces, and operators on those spaces are not introduced a priori,  in contrast to the usual discourse of quantum measurement theory\cite{holevo2011probabilistic,hayashi2006quantum}. Hence, all concepts related to physical measurements are defined in terms of classical information theory. Probabilistic aspects of measurements are introduced through random variables in the framework of probability theory. Solutions of a classical equation of motion, such as the equation for charged spinor fields, represent a classical charge distribution. The reason we restrict ourselves to classical theory is that some counterpart of the "quantum effect," i.e., uncertainty relations, already arises in classical field theory with no explicit quantization.
%
%
\section*{Physical Measurement}
We define terms associated with physical measurement according to classical estimation theory\cite{covert91-12-11} as follows. Let $\X$ be a random variable for a given physical system described by the $N$-tuple 
${\bm \theta}=\{\theta_1,\cdots,\theta_N\}$, where $\theta_i$ is the {\it i~th physical parameter}. The set of all possible values of 
$\theta_i \in \R$, denoted by $\Theta$, is called the {\it parameter set}. The random variable $\X$ is distributed according to the probability density function ${\rm f}(x;{\bm \theta})\geq 0$, which is normalized as $\int_{x\in \Omega} dx~{\rm f}(x;{\bm \theta})=1$, where $x\in\R$ is one possible value of the whole event $(=\Omega)$. 
For physical applications, we introduce the {\it probability amplitude} defined by
\begin{eqnarray*}
\left| 
\psi(x;{\bm \theta})
\right|^2={\rm f}(x;{\bm \theta}).
\end{eqnarray*}
It is assumed that the behavior of the physical system is determined by the probability amplitude rather than the probability density. This is analogous with quantum mechanical amplitude. However, the introduction of amplitude does not immediately mean that the theory has been quantized.
Considering probability density, the probability amplitude is only apparent by multiplication with a unitary operator, which implies that the amplitude contains more information than the density.\\
A part of {\it experimental apparatus} is assumed to output numbers distributed according to the probability density. Any resulting set of numbers $\X_n=\{x_1,\cdots,x_n \}$, drawn independently and identically distributed (i.i.d.), is called the {\it experimental data}. The estimate of the physical parameter is called a {\it measurement}. Because experimental data are i.i.d., the corresponding probability density function can be expressed as a product:
\begin{eqnarray*}
{\rm f}(\X_n;{\bm \theta})=\prod_{j=1}^n~{\rm f}(x_j;{\bm \theta}).
\end{eqnarray*}
A function mapping the experimental data to one possible value of the parameter set such as 
\begin{eqnarray*}
T_i:\X_n \rightarrow \Theta:\{x_1,\cdots,x_n\}\mapsto{\tilde \theta}_i
\end{eqnarray*}
is called an {\it estimator} for the $i$th physical parameter, denoted by
$
T_i(\X_n)={\tilde \theta_i}.
$
The {\it experimental error} in the $i$th physical parameter is defined as the root mean square error: 
\begin{eqnarray*}
\epsilon_i = E[(T_i(\X_n)-\theta_i)^2]^{1/2},
\end{eqnarray*}
where $\theta_i$ is the true value of the $i$ th physical parameter. True values of physical parameters are typically unknown, but a mean-square error can be reduced below any desired value by accumulating a sufficiently large amount of experimental data, thanks to the law of large numbers. If the mean value of the experimental error is zero, i.e.,
$
E_{\theta_i}[\tilde\theta_i-\theta_i]=0
$,
after accumulation of infinitely many statistics,
that estimator is called an {\it unbiased estimator}. Among such estimators, the one giving the least error is called the {\it best estimator}.\\
Simultaneous measurements of two physical parameters are described as follows. For the random variable $\X$, with two physical parameters ${\bm \theta}=\{\theta_1,\theta_2\}^T$, the experimental data with $n$ samplings can be expressed as
$
\X_n=\{{\bm x}_1,\cdots,{\bm x}_n\}=\{(x^1,x^2)_1^T,\cdots,(x^1,x^2)_n^T\}.
$
The probability density of the experimental data is expected to be a Gaussian distribution with two variables:
\begin{eqnarray*}
{\rm f}({\bm x_j};{\bm\theta})
=\frac{1}{2\pi |{\bm\sigma}|^{1/2}}
\exp
(
-({\bm x_j}-{\bm\theta})^T{\bm\sigma}^{-1}({\bm x_j}-{\bm\theta})/2,
),
\end{eqnarray*}
thanks to the central limit theorem. Here, ${\bm\sigma}$ is the covariance matrix of the data. When two parameters can be measured independently, the probability density function of the data becomes
\begin{eqnarray*}
{\rm f}\left({\bm x_j};{\bm\theta}\right)=\prod_{i=1,2}
\frac{1}{\sqrt{2\pi}\sigma_i^2}
\exp{\left(
\frac{-(x_j^i-\theta_i)^2}{2\sigma_i^2}
\right)}.
\end{eqnarray*}
In this case, it is known that the best estimators for physical parameters and experimental errors are given as follows:
\begin{eqnarray}
T_i(\X_n)&=&\frac{1}{n}\sum_{j=1}^n x_j^i=\tilde\theta_i,\nonumber\\
\epsilon_i(\X_n)^2&=&\frac{1}{n}\sum_{j=1}^n(x_j^i-\tilde\theta_i)^2=\tilde\sigma_i^2,\label{covdata}
\end{eqnarray}
for {i = 1, 2}.
%
%
\\
\indent
We now introduce quantitative informational properties. The $N$-tuple of random variables, 
\begin{eqnarray*}
V_i=\frac{\partial\log{{\rm f}(x,{\bm \theta})}}{\partial\theta_i},
\end{eqnarray*}
is called the {\it score}. 
It can be shown that the expectation value of the score is zero\cite{covert91-12-11}.
The score represents the sensitivity of the experimental data to the $i$th physical parameter. A large value of the score means that the experimental data are sensitive to the $i$th parameter and are expected to give a small error. 
The score for the experimental data can be shown to be
$
V_i(\X_n)=\sum_{j=1}^n V_i(x_j)
$
because the data are i.i.d. The covariance matrix of the score
\begin{eqnarray}
J_{ij}({\bm \theta})=\int dx~{\rm f}(x;{\bm \theta})
\frac{\partial\log{{\rm f}(x;{\bm \theta})}}{\partial\theta_i}
\frac{\partial\log{{\rm f}(x;{\bm \theta})}}{\partial\theta_j}
\end{eqnarray}
is called the {\it Fisher information matrix}~(FIM). FIM of the experimental data can be shown to be
$
J_{ij}(\X_n;{\bm \theta})=nJ_{ij}({\bm \theta})
$,
once again, because the data are i.i.d. ${\bm T}=\{T_1,\cdots,T_N\}$ is an unbiased estimator, and ${\bm \Sigma}({\bm \theta})$ is the covariance matrix of the data. Then, the Cram$\grave{\rm e}$r--Rao inequality
\begin{eqnarray}
{\bf \Sigma}({\bm \theta})\geq{\bm J}^{-1}({\bm \theta})\label{c-r}
\end{eqnarray}
holds true \cite{Cramer,Rao} as a matrix inequality, i.e., each element on the left-hand side is greater than or equal to each corresponding element on the right-hand side. This is one of the key items to consider when discussing uncertainty relations. Importance of the FIM and Cram$\grave{\rm e}$r--Rao inequality, and their application to the uncertainty relation, was also pointed out by Freiden\cite{Freiden,frieden2004science} and recently investigated by Watanabe et al.\cite{PhysRevA.84.042121}.
%
%
\section*{Equation of motion}
The equation of motion (EoM) is a differential equation describing the time evolution of a physical system. We assume that the physical system is governed by the probability amplitude, which is a solution of the EoM. Here, we assume the EoM satisfies the following conditions:~1)~it is a separable equation with respect to the time variable; ~2)~the spatial part after time-variable separation will be a holomorphic Sturm--Liouville function\cite{Hilbert}; and~3)~the Sturm--Liouville differential operator (SL operator) is self-adjoint. The SL-operator is defined by
\begin{eqnarray*}
L[\psi(x)]=\frac{d}{dx}\left(p(x)\frac{d}{dx}\psi(x)\right)
-q(x)\psi(x)+\lambda r(x)\psi(x)\label{s-leq},
\end{eqnarray*}
where $p(x), q(x)$, and $r(x)$ are smooth functions given by the dynamical system under consideration.
Solving the EoM leads to an eigenvalue problem of the form $L[\psi(x)]=0$ with the following boundary conditions on a finite interval $[a,b]$:
\begin{eqnarray*}
\psi(a)+\kappa_1\frac{d\psi(x)}{dx}|_{x=a}&=&0,\\
\psi(b)+\kappa_2\frac{d\psi(x)}{dx}|_{x=b}&=&0.
\end{eqnarray*}
Self-adjointness of the SL operator ensures that its eigenfunctions form a complete orthogonal system.
By using the complete orthogonal system obtained from the EoM, the generalized Fourier transformation (GFT) can be defined by
\begin{eqnarray}
{\hat f}(\lambda)=\int d\xi~f(\xi)\psi(\xi;\lambda), 
\end{eqnarray}
where $\psi(\lambda;\xi)$ is the eigenfunction of the EoM with eigenvalue $\lambda$. The existence of GFT with an appropriate integration measure and their inverse transformations are ensured by the generalized expansion theorem\cite{springerlink:10.1007/BF01474161,Tichmarsh,Kodaira1949,Kodaira1950}. For GFT, conservation of normalization of the two functions is ensured by the Parseval theorem\cite{kammler2007first}. This set of observables $\{\xi,\lambda\}$ is called the {\it GFT dual pair}. In the next section, we will obtain a nontrivial restriction on the uncertainty of a simultaneous measurement of the GFT dual pair.
%
%
\section*{Generalized uncertainty relations without disturbances}
\subsection*{Equation of wave motion}
The equation of wave motion is a typical example of an EoM as discussed in the last section. After separating out the time component, the steady-state solution in an infinitely large box is a plane wave solution of the form
\begin{eqnarray}
K(\xi_1;\xi_2)=\alpha_1 \exp{\left(\frac{i}{\tilde h}\xi_1\xi_2\right)}+
\alpha_2 \exp{\left(-\frac{i}{\tilde h}\xi_1\xi_2\right)}\label{wf},
\end{eqnarray}
where $\xi_i$ are two physical variables (observables) with appropriate dimensions and ${\tilde h}$ is a dimensional physical constant canceling out the dimensions of $\xi_1\xi_2$. It is well known that Eq. (\ref{wf}) specifies a complete orthogonal system, and the integral transformation with a kernel as given by Eq. (\ref{wf}) is the usual Fourier (inverse Fourier) transformation. Suppose that the experimental apparatus has been prepared such that the initial probability density for physical observables ($\xi_{1,2}$) is the Gaussian distribution
\begin{eqnarray}
{\rm f}(\xi_i,\{\mu_i,\sigma_i\})&=&
\frac{1}{\sqrt{2\pi}\sigma_i}\exp\left(
-\frac{(\xi_i-\mu_i)^2}{2\sigma_i^2}
\right),\label{gauss}
\end{eqnarray}
as we would naturally expect from the central limit theorem.
Then, we assume that these two observables constitute the GFT dual pair. In the wave function case, the GFT is equivalent to a usual Fourier transformation. 
Probability amplitudes describing the above probability densities are introduced as follows: 
\begin{eqnarray*}
&~&\psi_1(\xi_1;\{\bm\mu,\bm\sigma\})\\
&=&\sqrt{\frac{1}{\sqrt{2\pi}\sigma_1}}
\exp\left(
-\frac{(\xi_1-\mu_1)^2}{4\sigma_1^2}+i\frac{\mu_2(\xi_1-\mu_1)}{\tilde h}
\right).
\end{eqnarray*}
The Gaussian distribution of Eq. (\ref{gauss}) is obtained by squaring it.
After performing GFT, the probability amplitude of the GFT dual pair $\xi_2$ becomes 
\begin{eqnarray}
&~&\psi_2(\xi_2;\{\bm\mu,\bm\sigma\})\nonumber \\
&=&
\frac{1}{\sqrt{2\pi\tilde{h}}}
\int_{-\infty}^{\infty}d\xi_1~\psi_1(\xi_1;\{\bm\mu,\bm\sigma\})
\exp{\left(
-\frac{i}{\tilde h}\xi_1\xi_2
\right)}\nonumber \\
&=&\sqrt{\sqrt{\frac{2}{\pi}}\frac{\sigma_1}{\tilde h}}\exp\left(
-\frac{(\xi_2-\mu_2)^2\sigma_1^2}{\tilde{h}^2}-i\frac{\xi_2 \mu_1}{\tilde h}
\right).\label{amp2}
\end{eqnarray}
Because the transformation kernel in Eq. (\ref{wf}) is a general solution of the EoM as a function of the variable $\xi_1$, the GFT dual pair amplitude $\psi_2$ is the solution in the GFT dual space, by direct analogy with the relation between configuration and momentum spaces. By squaring the probability amplitude of Eq. (\ref{amp2}), one obtains the Gaussian distribution in the form
\begin{eqnarray*}
&~&{\rm f}(\xi_2;\{\mu_2,\sigma_2=\frac{\tilde h}{2\sigma_1}\})\\
&=&\sqrt{\frac{2}{\pi}}\frac{\sigma_1}{\tilde h}\exp\left(
-\frac{2(\xi_2-\mu_2)^2\sigma_1^2}{\tilde h^2}
\right),
\end{eqnarray*}
which has a standard deviation of $\sigma_2={\tilde h}/2\sigma_1$. Then, standard deviations of the two GFT dual parameters satisfy the relation
\begin{eqnarray}
\sigma_1 \sigma_2=\frac{\tilde h}{2}.\label{krunc}
\end{eqnarray}\\
\indent
Next, we show that this relation gives a lower bound on measurement errors. 
Preparation of the physical system with observable $\xi_1$ and measurement of the observable $\xi_2$ for the same system can be performed independently. This is justified because we are treating a classical field that has the wave equation as EoM. Then, that total probability density can be expressed as
$
{\rm f}({\bm\xi};\{{\bm\mu},{\bm\sigma}\})=
{\rm f}(\xi_1;\{\mu_1,\sigma_1\}){\rm f}(\xi_2;\{\mu_2,\sigma_2\}).
$
FIM can then be obtained from the above formula as follows: 
\begin{eqnarray}
&~&{\bm J}(\X_n;{\bm\mu})\nonumber\\
&=&\int_{-\infty}^\infty d\xi_1\int_{-\infty}^\infty d\xi_2~
{\rm f}({\bm\xi};{\bm\mu})
\frac{\partial\log{{\rm f}({\bm\xi};{\bm\mu})}}{\partial\mu_i}
\frac{\partial\log{{\rm f}({\bm\xi};{\bm\mu})}}{\partial\mu_j},\nonumber \\
&=&
\left(
\begin{array}{cc}
\frac{n}{\sigma_1^2}&0\\
0&\frac{n}{\sigma_2^2}
\end{array}
\right)=n{\bm J}_1({\bm\mu}).\label{fimdata}
\end{eqnarray}
Here we omit the $\sigma$-component of FIM, because it is not related to the uncertainty relation. On the other hand, the covariance matrix for simultaneous measurements of the two independent parameters is given as follows:
\begin{eqnarray}
{\bm \Sigma}(\bm\mu)&=&
\left(
\begin{array}{cc}
\frac{\sigma_1^2}{n}&0\\
0&\frac{\sigma_2^2}{n}
\end{array}
\right),\label{covmdata}
\end{eqnarray}
according to Eq. (\ref{covdata}). By comparing Eqs. (\ref{fimdata}) and (\ref{covmdata}), it can be observed that this experiment gives the lower limit of the Cram$\grave{\rm e}$r--Rao inequality [Eq. (\ref{c-r})]. Then, Eq. (\ref{krunc}) gives a lower bound on the initial distributions of the two parameters, and this in turn yields Kennard--Robertson type uncertainty relation,
$
\sigma_1\sigma_2\geq{\tilde h}/2.
$
%
%
\subsection*{Hydrogen atom}
As a second example, we consider the following EoM of the SL type operator\cite{landau1958quantum}.
\begin{eqnarray}
\frac{d^2}{dr^2}rR(r)+\frac{2}{a_{\small B}r} rR(r)-\kappa^2rR(r)=0.\label{hydeq}
\end{eqnarray}
This is the radial component of the Schr\"{o}dinger equation in a Coulomb potential, when angular momentum is zero. If we consider a hydrogen atom, $a_{\small B}=\hbar^2/m_ee^2=0.592\times10^{-8}$cm is called the $Bohr~radius$. It specifies a typical atomic length scale. Because we do not require any quantization, this is the equation for a classical electron field called the {\it de Broglie field}. Solutions of this equation give the charge density distribution of a classical electron field. A set of solutions normalized to unity in $0\leq r <\infty$ is given by 
\begin{eqnarray*}
R_n(r)=\sqrt{\left(\frac{2}{a_B n}\right)^3\frac{1}{2n^2}}\label{Lnorm}
\cdot\exp\left(-\frac{r}{a_B n}\right)
L^1_{n-1}\left(\frac{2r}{a_B n}\right),\label{hysol}
\end{eqnarray*}
where $L^\bullet_\bullet(\bullet)$ are Laguerre polynomials. This set consists of a complete orthonormal system satisfying
\begin{eqnarray}
\int_0^\infty dr~r^2R_n(r)R_m(r)=\delta_{nm}.\label{ors} 
\end{eqnarray}
The corresponding eigenvalues can be obtained as 
$
E_n=-{\hbar^2}\kappa^2/2m_e
$
with
$
\kappa=(a_Bn)^{-1}.
$
Once again we take the initial probability density of the electron to be a Gaussian distribution:
\begin{eqnarray}
\psi(r;\sigma_r)&=&
\left(\frac{1}{2\pi\sigma_r^2}\right)^{\frac{1}{4}}\exp\left(
-\frac{r^2}{4\sigma_r^2}
\right)\label{gbeam1},\\
{\rm f}(r;\sigma_r)&=&|\psi(r;\sigma_r)|^2,\nonumber \\
&=&\left(\frac{1}{2\pi\sigma_r^2}\right)^{\frac{1}{2}}\exp{\left(-\frac{r^2}{2\sigma_r^2}\right)}.\label{gbeam2}
\end{eqnarray}
One reason we start with the Gaussian distribution is that we expect it to be obtained from the central limit theorem for physical measurements of i.i.d. data and will lead to the minimum uncertainty condition. 
GFTs of these eigenfunctions are given as
\begin{eqnarray}
\phi_n(\sigma_r)&=&C(\sigma_r)\int_0^\infty dr~r^2\psi(r;\sigma_r)R_n(r),\label{Lgft}\\
&=&C(\sigma_r)\sum_{m=0}^{n-1}
\frac{(-1)^m2^{\frac{5}{4}+m}(m+2)\Gamma(n)}
{\pi^{\frac{1}{4}}m!\Gamma(n-m)(a_0n)^{\frac{3}{2}+m}}\nonumber \\
&\times&
F\left(\frac{m+3}{2},\frac{1}{2};\left(\frac{\sigma_r}{a_0n}\right)^2\right)
\sigma_r^{\frac{5}{2}+m},\label{phin}
\end{eqnarray}
where $F(\bullet,\bullet;\bullet)$ is the confluent hypergeometric series and $C(\sigma_r)$ is an appropriate normalization factor. 
Because the solutions $R_i(r)$ form a complete orthonormal set of functions, as shown in Eq. (\ref{ors}), the Gaussian distribution can be expressed as the inverse GFT of an infinite summation:
\begin{eqnarray*}
{\tilde \psi}(r;\sigma_r)&=&\sum_{i=1}^{\infty}\phi_i(\sigma_r)R_i(r).
\end{eqnarray*}
We have checked numerically that this inverse GFT will transfer $\phi_i(\sigma_r)$ back to the original Gaussian distribution, except it will be close to the origin. 
Note that atomic units (a.u.)\cite{landau1958quantum} are used in all numerical calculations in this study ($m_e=1,~\hbar=1,$ and $e=1$).
In this case, the usual Fourier transformation of $\psi(r;\sigma_r)$ has no clear physical meaning because the plane wave solution has no vanishing values at $r=\infty$, whereas $\psi(r\rightarrow\infty;\sigma_r)=0$ for a hydrogen atom. The GFT dual parameter of the radial coordinate $r$ must be an energy eigenvalue $E_n$. Probability density functions in the GFT dual space have Gaussian distributions, as shown in Figure~1. The relation between standard deviations of the initial Gaussian distribution and the corresponding energy spectrum is shown in Figure~2. We can observe that as the uncertainty of the energy spectrum decreases, position uncertainty increases. Moreover, products of these two uncertainties are almost constant at $\sigma_r\sigma_E\simeq 1/1.72$. This condition also gives the lower bound from the Cram$\grave{\rm e}$r--Rao inequality because the GFT dual probability density function ($\phi_n$) has a Gaussian distribution, and the discussion from the first example can still be applied. A generalized uncertainty relation for this case can be expressed as
\begin{eqnarray}
\sigma_r\cdot\sigma_E\geq \frac{a_B E_a}{1.72},
\end{eqnarray}
where $E_a$ is the atomic unit of energy. 
This relation is similar to the Kennard--Robertson uncertainty, except that the coefficient of the dimensional parameters on the right-hand side is $1/1.72$ instead of the usual value of $1/2$.\\
Let us look at this new relation from another point of view. The reason a hydrogen atom does not collapse to the size of the proton is usually discussed with only a qualitative reference to the uncertainty relation. Here, we investigate this matter in a quantitative way using the new uncertainty relation. The total energy of the electron can be given as
\begin{eqnarray*}
E_{H}=3\frac{p^2}{2}-\frac{1}{r}
\end{eqnarray*}
in a.u. The first and second terms are, basically, kinetic and potential energies, respectively. The factor of $3$ in front of kinetic energy originates from the spatial degree of freedom. Suppose now that the uncertainty relation is $\sigma_p\sigma_r\geq f_u$. In a ground state hydrogen atom, the electron is confined within the Bohr radius, $\sigma_r=a_B=r_0$, which suggests that $\sigma_p=f_u/\sigma_r=f_u/r_0$. The total energy can thus be written as
\begin{eqnarray*}
E^0_{H}=3\frac{f_u^2}{2r_0^2}-\frac{1}{r_0},
\end{eqnarray*}
where $r_0$ is the hydrogen radius in the ground state. The value $r_0$ must give the minimum energy; then, it can be obtained as a solution of 
$\partial E^0_{H}/\partial r_0=0$, 
i.e.,
$
r_0=3f_u^2.
$
If we set $r_0=a_B=1$, we get $f_u=1/\sqrt{3}=1/1.732...$, which is close to the factor in the new uncertainty relation. If one assumes that the uncertainty relation gives the exact hydrogen radius, then the uncertainty relation between GFT dual parameters for the system under the SL operator of Eq. (\ref{hydeq}) must be 
\begin{eqnarray}
\sigma_r\cdot\sigma_E\geq \frac{a_B E_a}{\sqrt{3}}.
\end{eqnarray}
We can give yet another view of the hydrogen atom in terms of classical information theory. The entropy of the energy spectrum of the hydrogen atom can be calculated numerically from Eq. (\ref{phin}) as
\begin{eqnarray*}
S(E)=-\sum_{n=1}^\infty \phi_n(\sigma_r)\log{\phi_n(\sigma_r)}.
\end{eqnarray*}
Numerical results are shown in Figure~3 as a function of the width of the radial distribution. The global minimum point of the entropy coincides with the Bohr radius, as shown Figure~3. 
%
%
\subsection*{Classical matter field in a cylinder}
We now consider a particle beam entering a cylindrical beam pipe. The Schr$\ddot{{\rm o}}$dinger equation in cylindrical coordinates is considered an EoM for the classical matter field. The EoM for a radial variable $R(r)$ with zero angular momentum can be expressed as
\begin{eqnarray*}
\frac{d}{dr}\left(r\frac{dR(r)}{dr}\right)-\frac{n^2}{r}R(r)
+(k^2-\lambda^2)rR(r)=0,
\end{eqnarray*}
which is also an SL-type equation. Solutions with the infinite potential barrier at $r=r_0$ are obtained as
\begin{eqnarray}
R_l^n(r)=\frac{\sqrt{2}}{r_0|J_{l+1}(z_l^n)|}J_l\left(z_l^n\frac{r}{r_0}\right),\label{Bessel}
\end{eqnarray}
where $J_l(x)$ is the $l$th order Bessel function of the first kind and 
$z_l^n$ is its $n$th zero point. This series of functions constitutes an orthonormal system with normalization given as
\begin{eqnarray*}
\int_0^\infty dr~r R_l^{n}(r)R_l^{m}(r)=\delta_{nm}.
\end{eqnarray*}
The corresponding eigenvalues are
\begin{eqnarray}
k^2&=&\left(\frac{z_l^n}{r_0}\right)^2+\lambda^2,\nonumber\\
P_l^n&=&\hbar~k\nonumber\\
&=&\hbar~\frac{z_l^n}{r_0}.\label{Bpv1}
\end{eqnarray}
Here, $\lambda$ is the eigenvalue from the appropriate boundary condition in the $z$-coordinate and will be neglected in the following discussion. Even though a physical constant $\hbar$ arises in the eigenvalues, this is a classical theory, and the solution $R_l^n(r)$ represents a classical {\it de Broglie field}. Suppose a particle beam with Gaussian distribution in the radial direction of the cylindrical coordinate enter a cylinder of radius $r_0$. The beam is assumed to be coaxial with respect to the cylinder, so the angular momentum of the beam is zero ($l=0$). This radial distribution can be expanded in terms of the solutions of Eq. (\ref{Bessel}) as
\begin{eqnarray*}
\phi_n(\sigma_R)=\int_0^{r_0}dr~r~\psi(r;\sigma_R)R_0^n(r),
\end{eqnarray*}
where $\psi(r;\sigma_R)$ is the radial distribution of the incident beam represented by Eq. (\ref{gbeam1}). When $\sigma_R\ll r_0$, the above integration can be performed analytically to give
\begin{eqnarray*}
\phi_n(\sigma_R)\approx\frac{4\sigma_R^{3/2}}
{(2\pi)^{1/4}r_0 J_1(z_0^n)}
\exp{\left(\frac{\sigma_R~z_0^n}{r_0}\right)^2}.
\end{eqnarray*}
The usual Fourier dual pair has no clear physical meaning because the solution of the EoM with a cylindrical boundary condition is not a plane wave. The GFT dual pair can be expressed as $\{r,~P_0^n=z_0^n/r_0\}$, and the GFT transforms the Gaussian to another Gaussian. The width of the $P_0^n$ distribution is proportional to the reciprocal of $\sigma_R$, as shown in Figure~4. The relation between the widths of the GFT pair is found numerically to be
$\sigma_R\sigma_P=\hbar/3.0$, which gives a smaller uncertainty than the Kennard--Robertson relation expected from the usual Fourier dual pair.
This reduction of the uncertainty can be explained with respect to the entropy. The entropy of the radial distribution, which is contained within the cylinder, can be calculated from the probability density function [Eq.(\ref{gbeam2})] as 
\begin{eqnarray*}
S^{\rm Gauss}(r<r_0;\sigma_R)
=-\int_{-r_0}^{r_0}dr~{\rm f}(r;\sigma_R)\log{{\rm f}(r;\sigma_R)},
\end{eqnarray*}
whereas the entropy of the momentum distribution can be obtained as
\begin{eqnarray*}
S^{\rm cyl}(p;\sigma_R)=-\sum_{n=1}^\infty\phi_n(\sigma_R)\log{\phi_n(\sigma_R)}.
\end{eqnarray*}
As beam width increases, entropy of the radial distribution increases; however, that of the momentum distribution decreases such that the sum of entropy and momentum is maintained constant, as shown in Figure~5. This behavior of entropies is easily understood qualitatively if we recall that entropy is a measure of the total amount of information in distributions. Under the minimum uncertainty condition, the total amount of information can be maintained constant. Here, let us consider the total amount of information quantitatively in order to estimate the information gain obtained from the boundary condition. When there are no boundary conditions on the radial distributions, the information entropy of the usual Fourier dual pair is 
\begin{eqnarray*}
S^{\rm Gauss}(\sigma_x)+S^{\rm Gauss}(\sigma_p)
&=&\frac{1}{2}\log{(4\pi^2 e^2\sigma_x^2\sigma_p^2)}+2\Delta,
\end{eqnarray*}
where $\Delta$ is an arbitrary value because of ambiguity in the choice of integration measure. Then, the relation between the entropy and the mean-square errors can be expressed as
\begin{eqnarray*}
\sigma_x\sigma_p&=&
\frac{1}{2\pi}\exp{\left(S(\sigma_x)+S(\sigma_p)-1-2\Delta\right)},\\
&=&\frac{\hbar}{2}\exp{\left(S(\sigma_x)+S(\sigma_p)\right)}.
\end{eqnarray*}
Here, we set $\Delta$ equal to $-1/2\log{(e\pi\hbar)}$ to obtain the minimum uncertainty relation when $S(\sigma_x)+S(\sigma_p)=0$. The average information due to the cylindrical boundary condition can be estimated as follows. The radial distribution of the incident beam is assumed to be the Gaussian distribution, as given in Eq. $(\ref{gauss})$. The total amount of information in the cylinder of radius $r_0$ can be calculated as
\begin{eqnarray*}
I(r_0,\sigma)&=&
\int_{-r_0}^{r_0}dr~{\rm f}(r,\{0,\sigma_R\})\\
&=&{\rm erf}\left(\frac{r_0}{\sqrt{2}\sigma_R}\right),
\end{eqnarray*}
where erf$(\bullet)$ is the error function. The entropy within the cylinder can be calculated as 
\begin{eqnarray*}
S^{\rm cyl}(r_0;\sigma_R)=-I(r_0,\sigma_R)\log{I(r_0,\sigma_R)}.
\end{eqnarray*}
The maximum value of the total information is found to be $S^{\rm cyl}=e^{-1}$ by solving $\partial S^{\rm cyl}/\partial r_0=0$.
This amount of information gain due to the cylindrical boundary condition can decrease the minimum uncertainty according to
\begin{eqnarray*}
\sigma_x\sigma_p&=&
\frac{\hbar}{2}\exp{\left(-\frac{1}{e}\right)},\\
&\simeq&\frac{\hbar}{2.9},
\end{eqnarray*}
which is consistent with the numerical result shown in Figure~4. This reduction of the uncertainty relation is due to information gain from the cylindrical boundary condition. 

%
%
\subsection*{Classical matter field in a sphere}
The last example is one of a classical matter field confined in a sphere. The calculations are almost the same as those in last the section; we list only the results here:
\begin{itemize}
\item EOM for the radial coordinate with zero angular momentum:
\begin{eqnarray*}
\frac{d}{dr}\left(r\frac{dR(r)}{dr}\right)+k^2rR(r)&=&0
\end{eqnarray*}
\item Solutions of EoM:
\begin{eqnarray*}
R_n(r)&=&\sqrt{\frac{2}{r_0}}\frac{\sin{(n\pi r/r_0)}}{r}
\end{eqnarray*}
\item Orthonormality relation:
\begin{eqnarray*}
\int_0^{r_0} dr~r^2 R_n(r)R_m(r)&=&\delta_{nm}
\end{eqnarray*}
\item Eigenvalue for the radial equation:
\begin{eqnarray*}
P_n&=&\hbar k\\
&=&\hbar\frac{n\pi}{r_0},~~(n=\pm 1,~\pm 2,\cdots)
\end{eqnarray*}
\item GFT and inverse GFT for the Gaussian distribution:
\begin{eqnarray*}
\phi_n(\sigma_R)&=&\int_0^{r_0}dr~r^2~\psi(r;\sigma_R)R_n(r)\\
\psi(r;\sigma_R)&=&\sum_{i=1}^{\infty}\phi_i(\sigma_R)R_i(r)
\end{eqnarray*}
\item GFT dual integration with approximation $\sigma_R\ll r_0$:
\begin{eqnarray*}
\phi_n(\sigma_R)&\simeq&\int_0^{\infty}dr~r^2~\psi(r;\sigma_R)R_n(r)\\
&=&\frac{2^{9/4}\pi^{5/4}}{r_0^{3/2}}
n\sigma_R^{5/2}\exp{\left(-\frac{n\pi\sigma_R}{r_0}\right)^2}
\end{eqnarray*}
\item Generalized uncertainty relation(see Figure~6):
\begin{eqnarray*}
\sigma_R\sigma_p&\simeq&\frac{\hbar}{2.9}\label{4158}
\end{eqnarray*}
\end{itemize}
In this case, mean-square errors once again give the Kennard--Robertson-type uncertainty relation with a factor of $\hbar/2.9$ instead of $\hbar/2$, which is consistent with the information gain because of the boundary condition, the same as in in the previous example.
\section*{Generalized uncertainty relations with disturbances}
\subsection*{The universally valid uncertainty relation}
Next we consider a disturbance because of the measurements themselves. 
It is assumed that the initial conditions of the physical system agree with the minimum uncertainty condition, with mean values of $\{\mu_1,\sigma_1^2\}$ and variances of $\{\mu_2,\sigma_2^2\}$. The two observables $\{\mu_1,\mu_2\}$ are assumed to be a GFT-pair. Estimators associated with this measurement are assumed to be unbiased. After simultaneous measurements of physical parameters $\{\mu_1,\mu_2\}$, final distributions are expected to be Gaussian because of the central limit theorem. The disturbance $\delta_i^2$ is defined as an increase in variance in the distribution of the $i$th observable after measurement. The convolution of two Gaussian distributions with mean values and variances as $\{\mu_1,\sigma_1^2\}$ and $\{\mu_2,\sigma_2^2\}$ yields the Gaussian distribution with $\{\mu_1,\tilde\sigma_1^2=\sigma_1^2+\delta_1^2\}$ and $\{\mu_2,\tilde\sigma_2^2=\sigma_2^2+\delta_2^2\}$. Initial and final distributions are not necessarily a GFT pair. To find the minimum uncertainty condition after measurement, we assume that measurement has been performed with very weak coupling and gives a minimum uncertainty pair of disturbances with $\delta_1\delta_2=\tilde{h}/2$. Then, standard deviations after measurement can be written in the form
\begin{eqnarray}
{\rm Min}[{\tilde \sigma}_1^2{\tilde \sigma}_2^2]
&=&
\left(\sigma_1^2+\delta_1^2\right)
\left(\sigma_2^2+\frac{\tilde{h}^2}{4\delta_1^2}\right).\label{580}
\end{eqnarray}
The minimum uncertainty condition after measurement is given by $\delta_1$ as $\partial{\rm Min}[{\tilde \sigma}_1^2{\tilde \sigma}_2^2]/\partial\delta_1=0$ such that 
\begin{eqnarray}
\frac{\partial{\rm Min}[{\tilde \sigma}_1^2{\tilde \sigma}_2^2]}
{\partial\delta_1}&=&
2\delta_1\sigma_2^2-\frac{\tilde{h}^2\sigma_1^2}{2\delta_1^3}=0\nonumber \\
&\Rightarrow&\sigma_1=\sqrt{\frac{\tilde{h}}{2}\frac{\sigma_1}{\sigma_2}}.\label{583}
\end{eqnarray}
Then, the formula
\begin{eqnarray}
{\tilde \sigma_1}{\tilde \sigma}_2\geq\sigma_1\sigma_2+\frac{\tilde{h}}{2}
\label{584}
\end{eqnarray}
follows from $(\ref{580})$ and $(\ref{583})$.
Under this condition, we derive the universally valid uncertainty relation (UVUR)\cite{Masanao200321,PhysRevA.67.042105,Masanao2004350}:
\begin{eqnarray}
\sigma_1\delta_2+\delta_1\sigma_2+\delta_1\delta_2\geq\frac{\tilde{h}}{2}\label{ozawa}.
\end{eqnarray}
From Eq. $(\ref{584})$ and the positivity of standard deviations $\sigma_i,\delta_i\geq0$, one can obtain 
\begin{eqnarray*}
(\sigma_1+\delta_1)^2(\sigma_2+\delta_2)^2
&\geq&(\sigma_1^2+\delta_1^2)(\sigma_2^2+\delta_2^2)\\
&=&\tilde\sigma_1^2\tilde\sigma_2^2\\
&\geq&\left(
\sigma_1\sigma_2+\frac{\tilde{h}}{2}
\right)^2.
\end{eqnarray*}
Then, the inequality
\begin{eqnarray*}
\sigma_1\sigma_2+\sigma_1\delta_2+\delta_1\sigma_2+\delta_1\delta_2&\geq&\sigma_1\sigma_2+\frac{\tilde{h}}{2}
\end{eqnarray*}
is obtained. The UVUR, Eq. $(\ref{ozawa})$, follows immediately from this formula\footnote{According to the proof of the UVUR in ref.\cite{Masanao200321}, the UVUR proposed must be {\it AND} of Eqs. (28) to (31) rather than the single inequality of Eq. (26) in \cite{Masanao200321}}.
\subsection*{Example: Hydrogen atom in a C$_{60}$ cage}
We now discuss the possibility of observing a violation of the lower bound in the Heisenberg-type uncertainty relation, $\delta_1\delta_2\geq\hbar/2$ (${\tilde h}$ is written as $\hbar$ in this section.). 

We reinvestigated position measurements of electrons in a hydrogen molecule trapped in a fullerene ${\rm C}_{60}$ cage. In 2005, Komatsu et al.~succeeded in encapsulating molecular hydrogen in fullerene ${\rm C}_{60}$ (${\rm H}_2@{\rm C}_{60}$) with an efficiency of approximately $100\%$\cite{Komatsu}. Sawa et al.~measured the electron density in the closed ${\rm C}_{60}$ cage using X-rays from a synchrotron light source. At first, the electron density in an open ${\rm C}_{60}$ cage was measured by using a BL-1A beamline at KEK\cite{Sawa1}. After Komatsu succeeded in encapsulating molecular hydrogen in a closed ${\rm C}_{60}$ cage, Sawa et al.~measured the electron-density again\cite{Sawa2}.
Electron density was measured using X-rays of wavelength $0.0998$~nm. The $H_2@{\rm C}_{60}$ sample was maintained at $50$K. It has been confirmed experimentally that the rotational mode of $H_2@{\rm C}_{60}$ is almost eliminated at this temperature\cite{Sawa3}. The inner diameter of fullerene ${\rm C}_{60}$ is known to be approximately $0.7$~nm. Here, we employ a radius of $0.36$~nm, as used in~\cite{Sawa1}. Electrons belonging to $H_2@{\rm C}_{60}$ can be described by the Schr{\" o}dinger equation with the boundary condition that electron density is zero for $r\geq0.36$~nm. The equation for the hydrogen atom has been solved numerically using $Mathematica$\cite{mathematica}. Here, we assume that energy eigenvalues are well approximated by those of the hydrogen atom for smaller principal quantum numbers. The solution proportional to $e^{+r}$, which is abandoned as a solution for molecular hydrogen in a vacuum, is allowed for H$_2$ in the ${\rm C}_{60}$ cage. Energy eigenvalues are listed in TABLE~\ref{ElevelC60}, along with the values in a vacuum. All solutions have zero angular momentum because there no solutions with nonzero angular momentum in a spherically symmetric cage of radius $0.36$~nm. Before measurement is taken, the electrons must be in the ground state. Uncertainty in the electron energy originates only from thermal fluctuations in electron energies, which is estimated to be $\sigma_E=4.31\times10^{-3}$~eV at a temperature of 50~K. The electron position in the ground state has a distribution with variance equal to the square of the Bohr radius. Hence, the Bohr radius is considered as the uncertainty in the initial electron position, i.e., $\sigma_r=a_B=5.29\times10^{-2}$~nm. The electron may be excited to one of the excited states listed in Table~1 \ref{ElevelC60} after the measurement. Then, the energy uncertainty of the electron is at most $\delta_E=27.24$~eV. The uncertainty in the electron density measurement is not clearly given, but the distribution of the electron density with respect to the radius has been measured very clearly, and the estimated number for the electrons in $H_2@{\rm C}_{60}$ is given as $1.9\pm0.4$\cite{Sawa1,Sawa2}. If we assume $\delta_r=1\times10^{-2}$~nm, the calculated electron density distribution is consistent with the figure and the estimated number of electrons in the cage is found to be $2.0\pm0.4$, which is consistent with the experimental result. We thus use the value $\delta_r=1\times10^{-2}$~nm for the position uncertainty of the measurement. In conclusion, we obtain the following uncertainties:
\begin{eqnarray*}
\delta_E\times\sigma r&\simeq&1.44~{\rm eV~nm},\\
\sigma_E\times\delta r&\simeq&4.31\times10^{-5}~{\rm eV~nm},\\
\delta_E\times\delta r&\simeq&2.72\times10^{-1}~{\rm eV~nm},\\
\frac{a_b E_0}{2}&\simeq&3.60\times10^{-1}~{\rm eV~nm}.
\end{eqnarray*}
Here, we observe that Heisenberg uncertainty pair $\delta_E\times\delta r$ exhibits smaller values than the expected vales $a_b E_0/2$.

%
%
\section*{Summary and conclusions}
We formulated the physical measurement process on the basis of classical information theory without introducing any quantization of the physical system. The probabilistic behavior of physical measurements arises from the assumption that the experimental data are random variables obeying a probability law. Even in this classical context, essential properties of uncertainty relations are exhibited. The Kennard--Robertson type of uncertainty relation arises for two physical quantities related to one another by Fourier transformation. Although Fourier transformation plays an essential role, it is not the only transformation leading to an uncertainty relation. We investigated GFT dual pairs of physical quantities governed by Sturm--Liouville-type differential equations and obtained a generalized uncertainty relation for such GFT dual pairs. In addition, we showed that Gaussian distributions realize the minimum uncertainty condition using the Cram$\grave{\rm e}$r--Rao inequality, and the minimum uncertainty condition can give a smaller error than that implied by the usual Kennard--Robertson lower limit. This reduction of uncertainties can be understood quantitatively in terms of information entropy because of boundary conditions.
\section*{Acknowledgments}
  \ifthenelse{\boolean{publ}}{\small}{}
 We wish to thank Dr. Y.~Sugiyama for his continuous encouragement and fruitful discussions and Prof. Tsutsui for his excellent lecture about advanced quantum mechanics. We would like to thank Enago for the English language review.

\newpage
{\ifthenelse{\boolean{publ}}{\footnotesize}{\small}
 \bibliographystyle{bmc_article}  
  \bibliography{meastheorunif_ER2} }     


\begin{thebibliography}{10}
\providecommand{\url}[1]{[#1]}
\providecommand{\urlprefix}{}

\bibitem{Heisenberg}
Heisenberg W: \emph{Z. Phys.} 1927, \textbf{43}:172.

\bibitem{Kennard}
Kennard EH: \emph{Z. Phys.} 1927, \textbf{44}:326.

\bibitem{Robertson}
Robertson H: \emph{Phys. Rev.} 1929, \textbf{34}:163.

\bibitem{Ozawa1988zz}
Ozawa M: \emph{Phys. Rev. Lett.} 1988, \textbf{60}:385.

\bibitem{Ishikawa}
Ishikawa S: \emph{Rep. Math. Phys.} 1991, \textbf{29}:257.

\bibitem{Masanao200321}
Ozawa M: \emph{Physics Letters A} 2003, \textbf{318}:21.

\bibitem{PhysRevA.67.042105}
Ozawa M: \emph{Phys. Rev. A} 2003, \textbf{67}:042105.

\bibitem{Masanao2004350}
Ozawa M: \emph{Ann. Phys.} 2004, \textbf{311}(2):350.

\bibitem{10228899}
Erhart J, Sponar S, Sulyok G, Badurek G, Ozawa M, Hasegawa Y: \emph{Nat. Phys.,
  Jan. 2012} 2012, \textbf{advance online publication}.

\bibitem{Kurihara:arXiv1201.5151}
Kurihara Y: \textbf{Comment on "Experimental demonstration of a universally
  valid error-disturbance uncertainty relation in spin measurements"} 2012.

\bibitem{holevo2011probabilistic}
Holevo A: \emph{Probabilistic and Statistical Aspects of Quantum Theory}.
  Quaderni Monographs, Edizioni Della Normale 2011.

\bibitem{hayashi2006quantum}
Hayashi M: \emph{Quantum information: an introduction}. Springer 2006.

\bibitem{covert91-12-11}
Cover T, Thomas J: \emph{Elements of information theory}. New York: Wiley 1991.

\bibitem{Cramer}
Cram$\grave{\rm e}$r H: \emph{Mathematical Method of Statisticas}. Princeton
  University Press 1946.

\bibitem{Rao}
Rao CR: \textbf{Information and accuracy obtainable in the estimation of
  statistical parameters}. \emph{Bull. Calcutta Math. Soc.} 1945,
  \textbf{37}:81.

\bibitem{Freiden}
Freiden BR: \emph{Phys. Lett. A} 1992, \textbf{169}:123.

\bibitem{frieden2004science}
Freiden BR: \emph{Science from Fisher information: a unification}. Cambridge
  University Press 2004.

\bibitem{PhysRevA.84.042121}
Watanabe Y, Sagawa T, Ueda M: \emph{Phys. Rev. A} 2011, \textbf{84}:042121.

\bibitem{Hilbert}
Courant R, Hilbert D: \emph{Methods of Mathematical Physics}, Wiley-VCH (1924),
  (1989) :291 -- 295.

\bibitem{springerlink:10.1007/BF01474161}
Weyl H: \textbf{$\ddot{{\rm U}}$ber gew$\ddot{{\rm o}}$hnliche
  Differentialgleichungen mit Singularit$\ddot{{\rm a}}$ten und die
  zugeh$\ddot{{\rm o}}$rigen Entwicklungen willk$\ddot{{\rm u}}$rlicher
  Funktionen}. \emph{Mathematische Annalen} 1910, \textbf{68}:220 -- 269.

\bibitem{Tichmarsh}
Tichmarsh EC: \emph{Eigenfunction expansions associated with second order
  differential equations}. Oxford University Press 1946.

\bibitem{Kodaira1949}
Kodaira K: \emph{Amer. J. Math.} 1949, \textbf{71}:921.

\bibitem{Kodaira1950}
Kodaira K: \emph{Amer. J. Math.} 1950, \textbf{72}:502.

\bibitem{kammler2007first}
Kammler D: \emph{A first course in fourier analysis}. Cambridge University
  Press 2007.

\bibitem{landau1958quantum}
Landau L, Lifshitz E: \emph{Quantum mechanics, non-relativistic theory}. A-W
  series in advanced physics, Pergamon Press 1958.

\bibitem{Komatsu}
Komatsu K, Murata M, Murata Y: \emph{Science} 2005, \textbf{307}:238.

\bibitem{Sawa1}
Sawa H, Wakabayashi Y, Murata Y, Murata M, Komatsu K: \emph{Angew. Chem. int.
  ed} 2005, \textbf{44}:1981.

\bibitem{Sawa2}
Sawa H: . [Private communication].

\bibitem{Sawa3}
Kohama Y, Rachi T, Jing J, Li Z, Tang J, Kumashiro R, Izumisawa S, Kawaji T H
  nd~Atake, Sawa H, Murata Y, Komatsu K, Tanigaki K: \emph{Phys. Rev. Lett.}
  2009, \textbf{103}:073001.

\bibitem{mathematica}
\emph{Mathematica Edition: Version 8.0}. Wolfram Research, Inc. 2010.

\end{thebibliography}

\newcommand{\BMCxmlcomment}[1]{}

\BMCxmlcomment{

<refgrp>

<bibl id="B1">
  <aug>
    <au><snm>Heisenberg</snm><fnm>W.</fnm></au>
  </aug>
  <source>Z. Phys.</source>
  <pubdate>1927</pubdate>
  <volume>43</volume>
  <fpage>172</fpage>
</bibl>

<bibl id="B2">
  <aug>
    <au><snm>Kennard</snm><fnm>E. H.</fnm></au>
  </aug>
  <source>Z. Phys.</source>
  <pubdate>1927</pubdate>
  <volume>44</volume>
  <fpage>326</fpage>
</bibl>

<bibl id="B3">
  <aug>
    <au><snm>Robertson</snm><fnm>H.P.</fnm></au>
  </aug>
  <source>Phys. Rev.</source>
  <pubdate>1929</pubdate>
  <volume>34</volume>
  <fpage>163</fpage>
</bibl>

<bibl id="B4">
  <aug>
    <au><snm>Ozawa</snm><fnm>M</fnm></au>
  </aug>
  <source>Phys. Rev. Lett.</source>
  <pubdate>1988</pubdate>
  <volume>60</volume>
  <fpage>385</fpage>
</bibl>

<bibl id="B5">
  <aug>
    <au><snm>Ishikawa</snm><fnm>S.</fnm></au>
  </aug>
  <source>Rep. Math. Phys.</source>
  <pubdate>1991</pubdate>
  <volume>29</volume>
  <fpage>257</fpage>
</bibl>

<bibl id="B6">
  <aug>
    <au><snm>Ozawa</snm><fnm>M</fnm></au>
  </aug>
  <source>Physics Letters A</source>
  <pubdate>2003</pubdate>
  <volume>318</volume>
  <fpage>21</fpage>
</bibl>

<bibl id="B7">
  <aug>
    <au><snm>Ozawa</snm><fnm>M</fnm></au>
  </aug>
  <source>Phys. Rev. A</source>
  <pubdate>2003</pubdate>
  <volume>67</volume>
  <fpage>042105</fpage>
</bibl>

<bibl id="B8">
  <aug>
    <au><snm>Ozawa</snm><fnm>M</fnm></au>
  </aug>
  <source>Ann. Phys.</source>
  <pubdate>2004</pubdate>
  <volume>311</volume>
  <issue>2</issue>
  <fpage>350</fpage>
</bibl>

<bibl id="B9">
  <aug>
    <au><snm>Erhart</snm><fnm>J</fnm></au>
    <au><snm>Sponar</snm><fnm>S</fnm></au>
    <au><snm>Sulyok</snm><fnm>G</fnm></au>
    <au><snm>Badurek</snm><fnm>G</fnm></au>
    <au><snm>Ozawa</snm><fnm>M</fnm></au>
    <au><snm>Hasegawa</snm><fnm>Y</fnm></au>
  </aug>
  <source>Nat. Phys., Jan. 2012</source>
  <publisher>Nature Publishing Group</publisher>
  <pubdate>2012</pubdate>
  <volume>advance online publication</volume>
</bibl>

<bibl id="B10">
  <title><p>Comment on "Experimental demonstration of a universally valid
  error-disturbance uncertainty relation in spin measurements"</p></title>
  <aug>
    <au><snm>Kurihara</snm><fnm>Y</fnm></au>
  </aug>
  <pubdate>2012</pubdate>
</bibl>

<bibl id="B11">
  <title><p>Probabilistic and Statistical Aspects of Quantum Theory</p></title>
  <aug>
    <au><snm>Holevo</snm><fnm>A.S.</fnm></au>
  </aug>
  <publisher>Edizioni Della Normale</publisher>
  <series><title><p>Quaderni Monographs</p></title></series>
  <pubdate>2011</pubdate>
</bibl>

<bibl id="B12">
  <title><p>Quantum information: an introduction</p></title>
  <aug>
    <au><snm>Hayashi</snm><fnm>M.</fnm></au>
  </aug>
  <publisher>Springer</publisher>
  <pubdate>2006</pubdate>
</bibl>

<bibl id="B13">
  <title><p>Elements of information theory</p></title>
  <aug>
    <au><snm>Cover</snm><fnm>T.</fnm></au>
    <au><snm>Thomas</snm><fnm>J.</fnm></au>
  </aug>
  <publisher>New York: Wiley</publisher>
  <pubdate>1991</pubdate>
</bibl>

<bibl id="B14">
  <title><p>Mathematical Method of Statisticas</p></title>
  <aug>
    <au><snm>Cram$\grave{\rm e}$r</snm><fnm>H</fnm></au>
  </aug>
  <publisher>Princeton University Press</publisher>
  <pubdate>1946</pubdate>
</bibl>

<bibl id="B15">
  <title><p>Information and accuracy obtainable in the estimation of
  statistical parameters</p></title>
  <aug>
    <au><snm>Rao</snm><fnm>C. R.</fnm></au>
  </aug>
  <source>Bull. Calcutta Math. Soc.</source>
  <pubdate>1945</pubdate>
  <volume>37</volume>
  <fpage>81</fpage>
</bibl>

<bibl id="B16">
  <aug>
    <au><snm>Freiden</snm><fnm>BR</fnm></au>
  </aug>
  <source>Phys. Lett. A</source>
  <pubdate>1992</pubdate>
  <volume>169</volume>
  <fpage>123</fpage>
</bibl>

<bibl id="B17">
  <title><p>Science from Fisher information: a unification</p></title>
  <aug>
    <au><snm>Freiden</snm><fnm>BR</fnm></au>
  </aug>
  <publisher>Cambridge University Press</publisher>
  <pubdate>2004</pubdate>
</bibl>

<bibl id="B18">
  <aug>
    <au><snm>Watanabe</snm><fnm>Y</fnm></au>
    <au><snm>Sagawa</snm><fnm>T</fnm></au>
    <au><snm>Ueda</snm><fnm>M</fnm></au>
  </aug>
  <source>Phys. Rev. A</source>
  <pubdate>2011</pubdate>
  <volume>84</volume>
  <fpage>042121</fpage>
</bibl>

<bibl id="B19">
  <title><p>Methods of Mathematical Physics</p></title>
  <aug>
    <au><snm>Courant</snm><fnm>R</fnm></au>
    <au><snm>Hilbert</snm><fnm>D</fnm></au>
  </aug>
  <publisher>Wiley-VCH</publisher>
  <pubdate>(1924), (1989)</pubdate>
  <fpage>291</fpage>
  <lpage>-295</lpage>
</bibl>

<bibl id="B20">
  <title><p>$\ddot{{\rm U}}$ber gew$\ddot{{\rm o}}$hnliche
  Differentialgleichungen mit Singularit$\ddot{{\rm a}}$ten und die
  zugeh$\ddot{{\rm o}}$rigen Entwicklungen willk$\ddot{{\rm u}}$rlicher
  Funktionen</p></title>
  <aug>
    <au><snm>Weyl</snm><fnm>H</fnm></au>
  </aug>
  <source>Mathematische Annalen</source>
  <publisher>Springer Berlin / Heidelberg</publisher>
  <pubdate>1910</pubdate>
  <volume>68</volume>
  <fpage>220</fpage>
  <lpage>-269</lpage>
</bibl>

<bibl id="B21">
  <title><p>Eigenfunction expansions associated with second order differential
  equations</p></title>
  <aug>
    <au><snm>Tichmarsh</snm><fnm>EC</fnm></au>
  </aug>
  <publisher>Oxford University Press</publisher>
  <pubdate>1946</pubdate>
</bibl>

<bibl id="B22">
  <aug>
    <au><snm>Kodaira</snm><fnm>K</fnm></au>
  </aug>
  <source>Amer. J. Math.</source>
  <publisher>The Johns Hopkins University Press</publisher>
  <pubdate>1949</pubdate>
  <volume>71</volume>
  <fpage>921</fpage>
</bibl>

<bibl id="B23">
  <aug>
    <au><snm>Kodaira</snm><fnm>K</fnm></au>
  </aug>
  <source>Amer. J. Math.</source>
  <publisher>The Johns Hopkins University Press</publisher>
  <pubdate>1950</pubdate>
  <volume>72</volume>
  <fpage>502</fpage>
</bibl>

<bibl id="B24">
  <title><p>A first course in fourier analysis</p></title>
  <aug>
    <au><snm>Kammler</snm><fnm>D.W.</fnm></au>
  </aug>
  <publisher>Cambridge University Press</publisher>
  <pubdate>2007</pubdate>
  <fpage>73</fpage>
  <lpage>-74</lpage>
</bibl>

<bibl id="B25">
  <title><p>Quantum mechanics, non-relativistic theory</p></title>
  <aug>
    <au><snm>Landau</snm><fnm>L.D.</fnm></au>
    <au><snm>Lifshitz</snm><fnm>E.M.</fnm></au>
  </aug>
  <publisher>Pergamon Press</publisher>
  <series><title><p>A-W series in advanced physics</p></title></series>
  <pubdate>1958</pubdate>
</bibl>

<bibl id="B26">
  <aug>
    <au><snm>Komatsu</snm><fnm>K</fnm></au>
    <au><snm>Murata</snm><fnm>M</fnm></au>
    <au><snm>Murata</snm><fnm>Y</fnm></au>
  </aug>
  <source>Science</source>
  <pubdate>2005</pubdate>
  <volume>307</volume>
  <fpage>238</fpage>
</bibl>

<bibl id="B27">
  <aug>
    <au><snm>Sawa</snm><fnm>H</fnm></au>
    <au><snm>Wakabayashi</snm><fnm>Y</fnm></au>
    <au><snm>Murata</snm><fnm>Y</fnm></au>
    <au><snm>Murata</snm><fnm>M</fnm></au>
    <au><snm>Komatsu</snm><fnm>K</fnm></au>
  </aug>
  <source>Angew. Chem. int. ed</source>
  <pubdate>2005</pubdate>
  <volume>44</volume>
  <fpage>1981</fpage>
</bibl>

<bibl id="B28">
  <aug>
    <au><snm>Sawa</snm><fnm>H</fnm></au>
  </aug>
  <note>Private communication</note>
</bibl>

<bibl id="B29">
  <aug>
    <au><snm>Kohama</snm><fnm>Y.</fnm></au>
    <au><snm>Rachi</snm><fnm>T.</fnm></au>
    <au><snm>Jing</snm><fnm>J.</fnm></au>
    <au><snm>Li</snm><fnm>Z.</fnm></au>
    <au><snm>Tang</snm><fnm>J.</fnm></au>
    <au><snm>Kumashiro</snm><fnm>R.</fnm></au>
    <au><snm>Izumisawa</snm><fnm>S.</fnm></au>
    <au><snm>Kawaji</snm><fnm>T.</fnm></au>
    <au><snm>Sawa</snm><fnm>H.</fnm></au>
    <au><snm>Murata</snm><fnm>Y.</fnm></au>
    <au><snm>Komatsu</snm><fnm>K.</fnm></au>
    <au><snm>Tanigaki</snm><fnm>K.</fnm></au>
  </aug>
  <source>Phys. Rev. Lett.</source>
  <pubdate>2009</pubdate>
  <volume>103</volume>
  <fpage>073001</fpage>
</bibl>

<bibl id="B30">
  <title><p>Mathematica Edition: Version 8.0</p></title>
  <publisher>Wolfram Research, Inc.</publisher>
  <pubdate>2010</pubdate>
</bibl>

</refgrp>
} 


\ifthenelse{\boolean{publ}}{\end{multicols}}{}
\section*{Figures}
  \subsection*{Figure 1 - The probability density function}\label{Hy1}
Example of probability density function in GFT dual space for radial distribution of classical electron field in a hydrogen atom. The solid line is a Gaussian fit. A Gaussian distribution is assumed for radial variable, with variance
$\sigma_r^2=10^2$.
\begin{figure}[h]
\begin{center}
 \epsfig{file=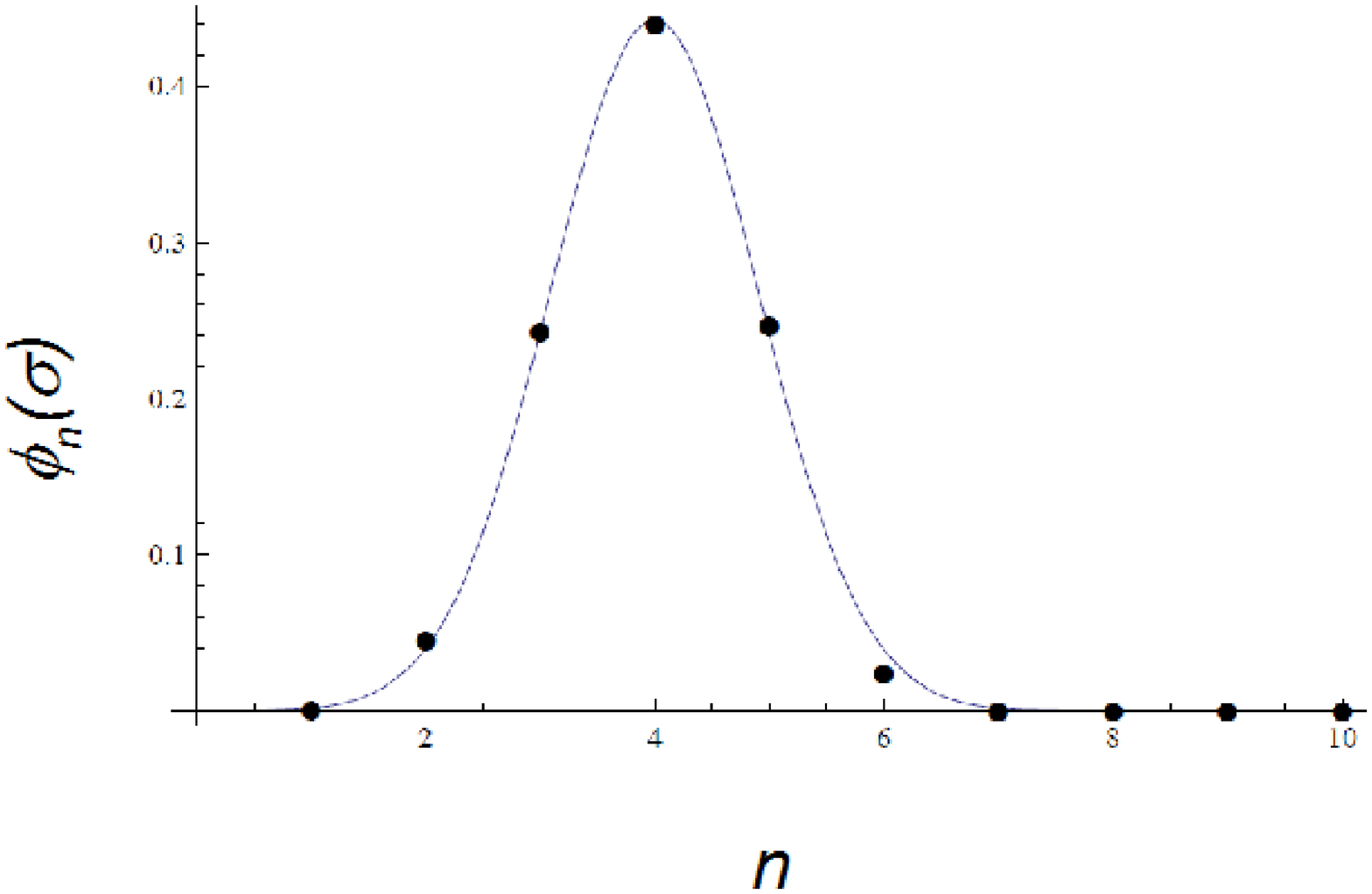,height=4cm}
\end{center}
\end{figure}
  \subsection*{Figure 2 - Relation between standard deviations of initial Gaussian distribution and corresponding energy spectrum}\label{Hy2}
Relation between standard deviations of initial Gaussian distribution and corresponding energy spectrum. Electron mass and Bohr radius are set to unity (a.u.) in calculations.
\begin{figure}[h]
\begin{center}
 \epsfig{file=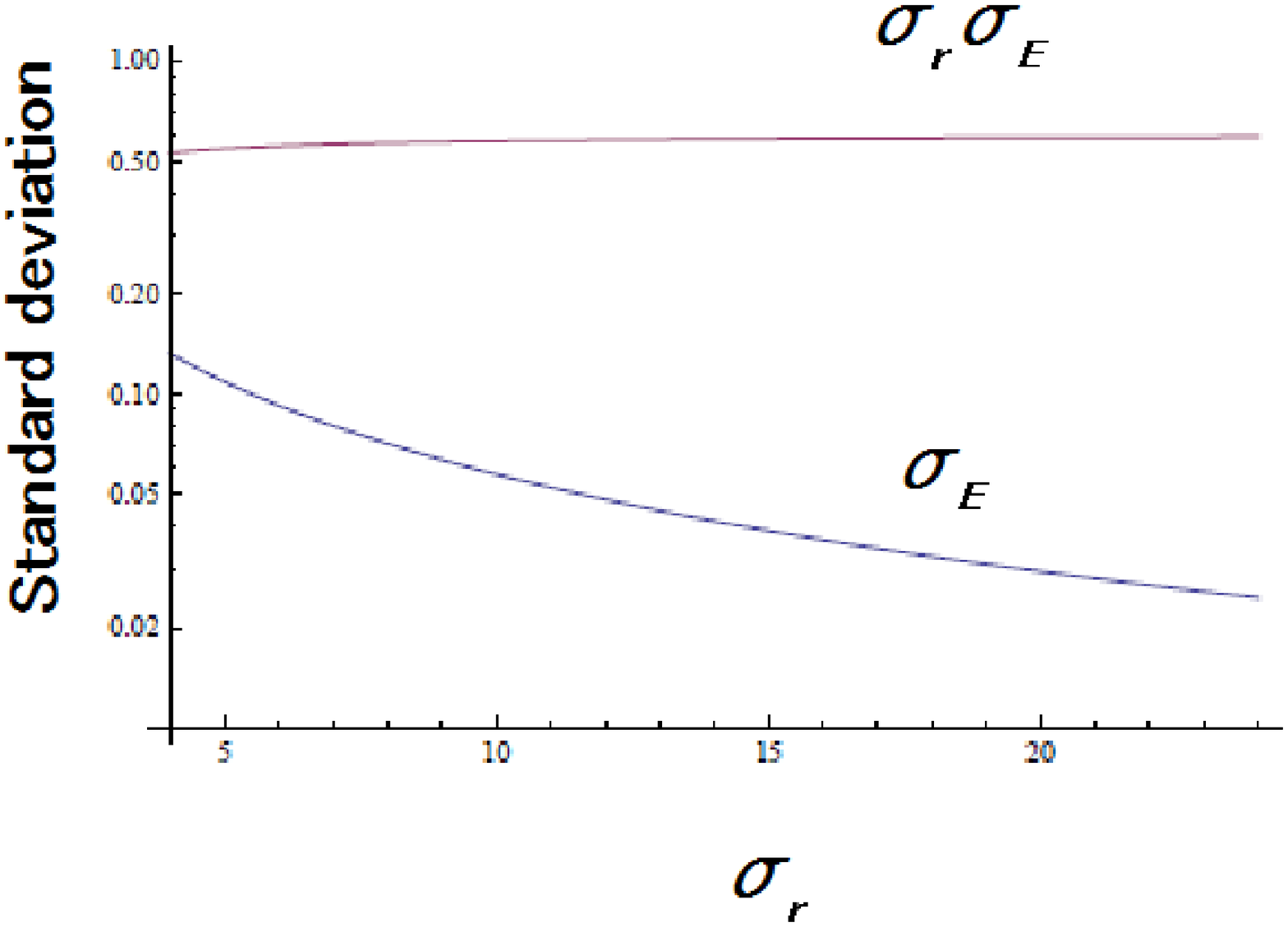,height=4cm}
\end{center}
\end{figure}
  \subsection*{Figure 3 - Entropy obtained from hydrogen energy states and radial distributions}\label{Hy3}
Entropy obtained from hydrogen energy states and radial distributions as function of width of radial distribution ($\sigma_r$). Entropy of hydrogen energy states reaches minimum at Bohr radius.
\begin{figure}[h]
\begin{center}
 \epsfig{file=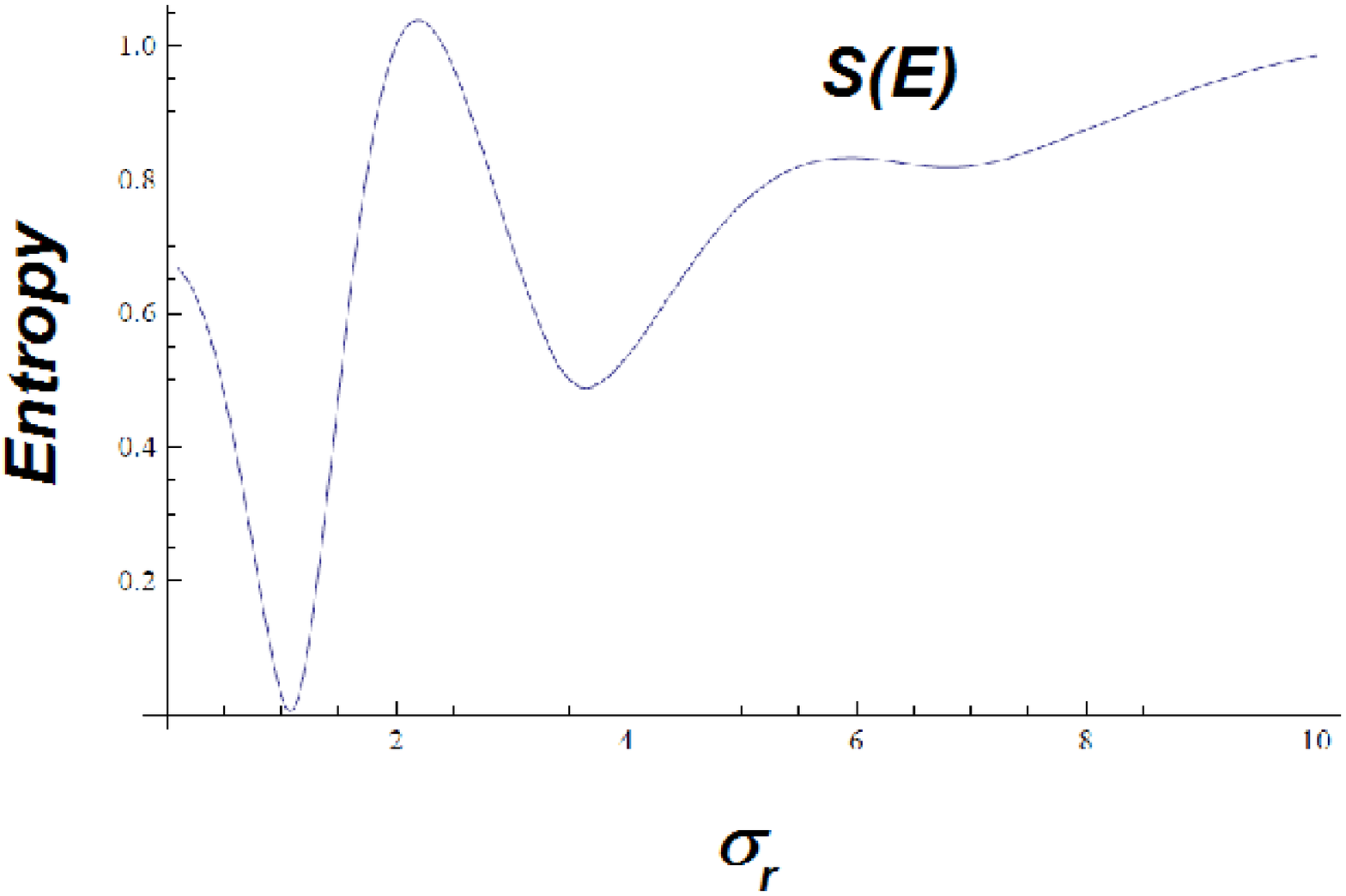,height=4cm}
\end{center}
\end{figure}
  \subsection*{Figure 4 - Width of radial distribution of beam}\label{cyl1}
Width of radial distribution of beam multiplied by its momentum as function of beam width. GFT can maintain product of widths of pair at $\sigma_R\sigma_P=1/0.30$, where $\hbar$ is set to unity.
\begin{figure}[h]
\begin{center}
 \epsfig{file=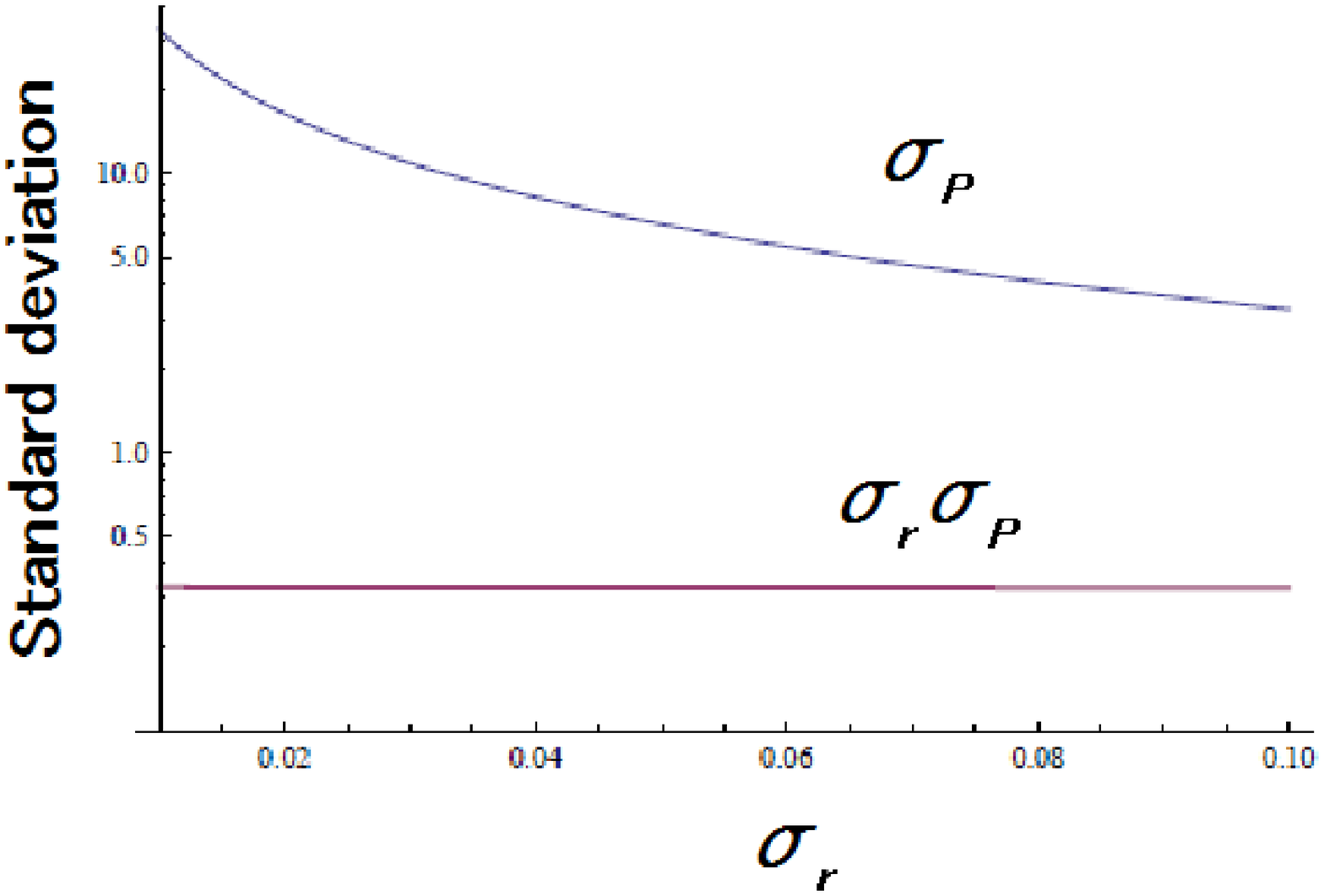,height=4cm}
\end{center}
\end{figure}
  \subsection*{Figure 5 - Entropies of radial distribution of beam}\label{cyl2}
Entropies of radial distribution of beam and those of GFT dual pair of momentum eigenstate distribution. $\hbar$ and $r_0$ are set to unity.
\begin{figure}[h]
\begin{center}
 \epsfig{file=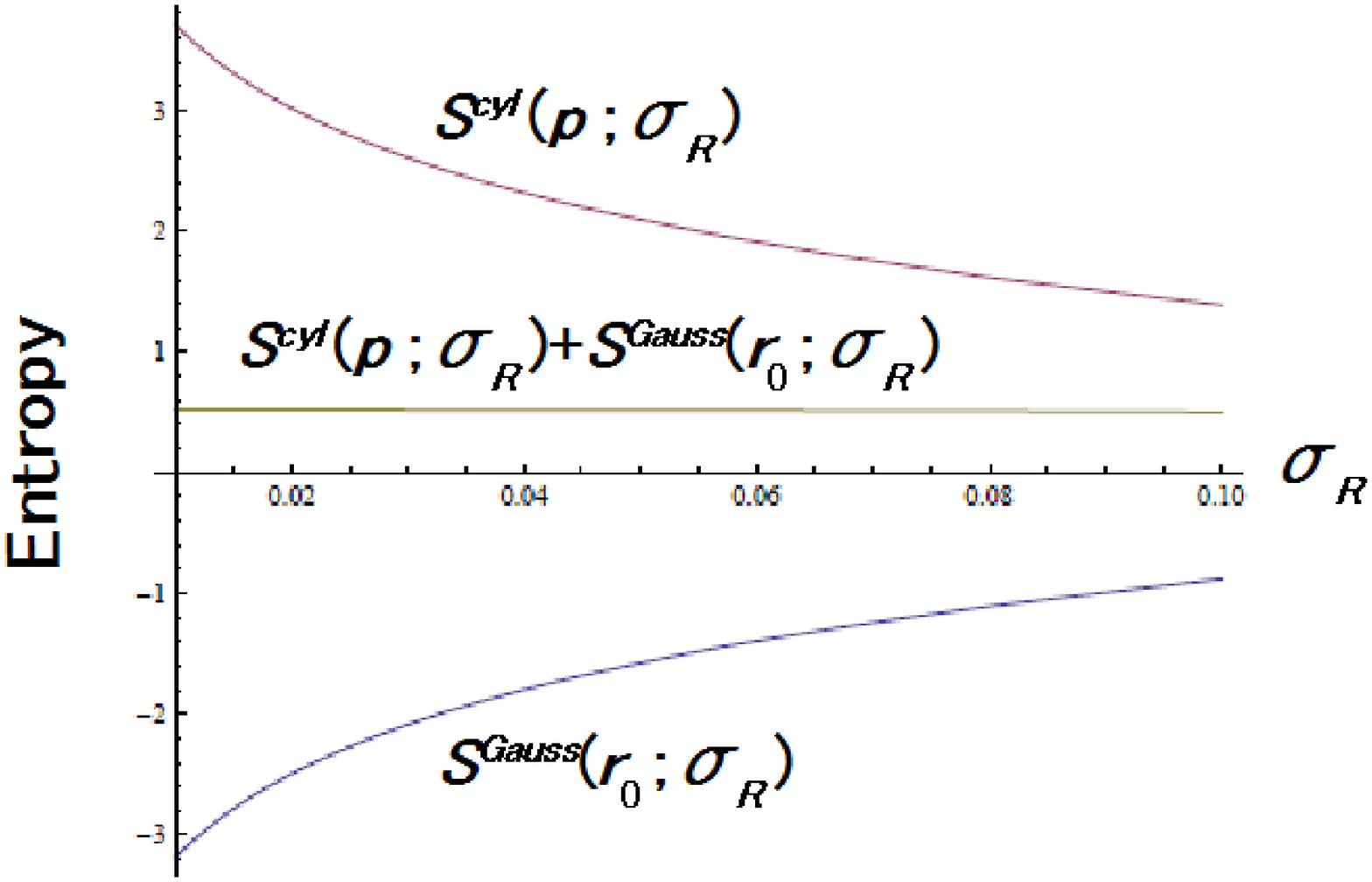,height=4cm}
\end{center}
\end{figure}
  \subsection*{Figure 6 - Width of radial distribution of initial field}\label{sphere1}
Width of radial distribution of initial field multiplied by its GFT dual momentum. GFT can maintain product of widths of pair at $\sigma_r\sigma_P=1/2.9$, where $\hbar$ is set to unity.
\begin{figure}[h]
\begin{center}
 \epsfig{file=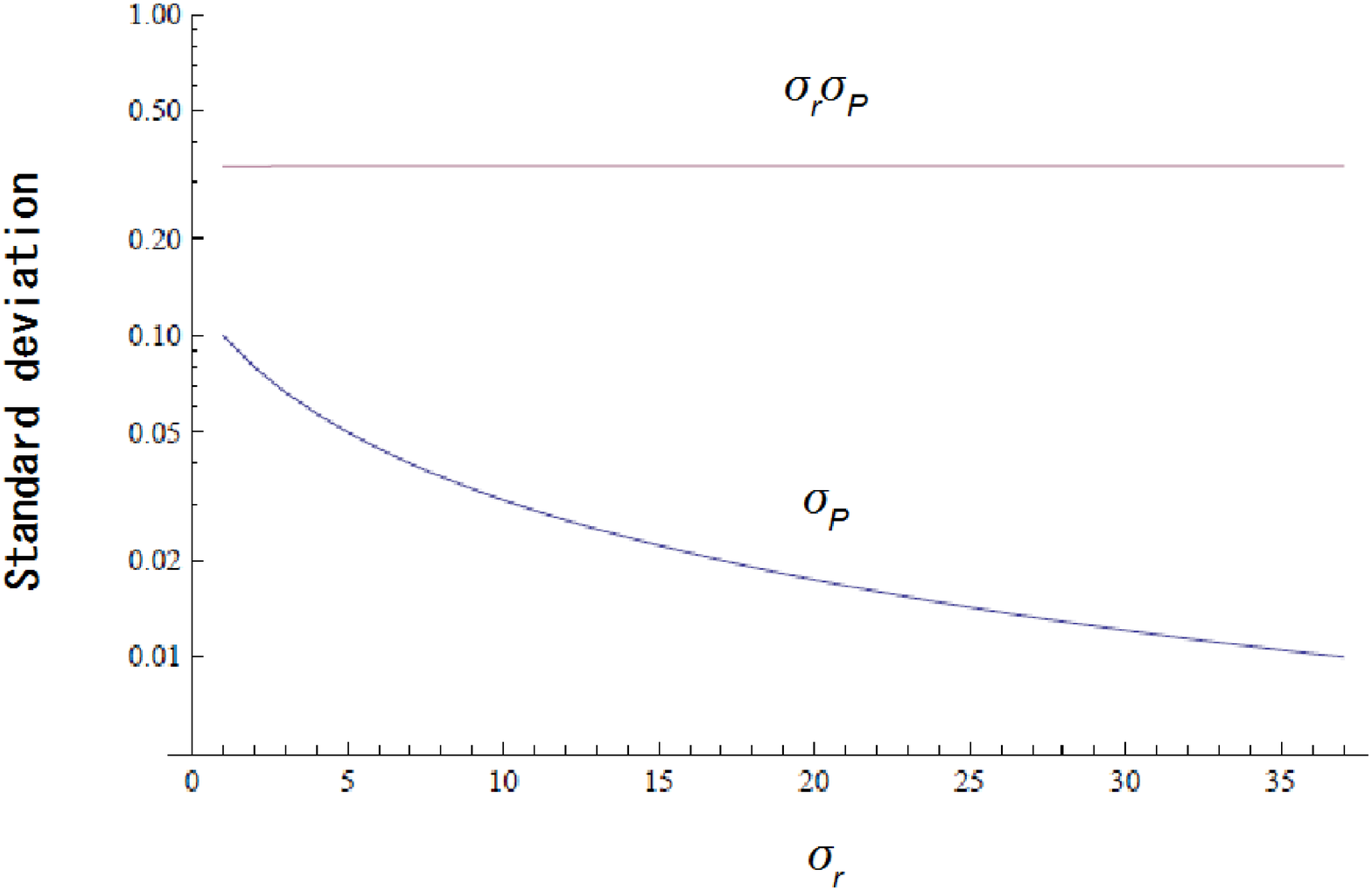,height=4cm}
\end{center}
\end{figure}
\section*{Tables}
  \subsection*{Table 1 - Energy eigenvalues of hydrogen atom with boundary condition}\label{ElevelC60}
Energy eigenvalues of hydrogen atom with boundary condition from ${\rm C}_{60}$ cage. Here, $n$ is the principal quantum number. In ${\rm C}_{60}$ cage, solutions proportional to $e^{+r}$ (Sol.2) are permitted, as are those proportional to $e^{-r}$ (Sol.1), in contrast with the situation in a vacuum (Vac.).\par \mbox{}

    \par
    \mbox{
  \begin{tabular}{|c||r|r|r|}
\hline
$n$&H2(Vac.)&$H_2@{\rm C}_{60}$(Sol.$1$)&$H_2@{\rm C}_{60}$(Sol.$2$)\\ \hline
1&-13.62&-13.62&-13.62\\
2&-3.41&-1.14&-1.14\\
3&-1.51&-&1.14\\
4&-0.85&-&13.62\\
\hline
  \end{tabular}
      }
\end{bmcformat}
\end{document}